\documentclass[12pt]{iopart}

\usepackage{graphicx}
\usepackage[dvipsnames]{xcolor}

\begin{document}

\title[Collective Network Coloring Problems]{Random Choices can Facilitate the Solving of Collective Network Coloring Problems by Artificial Agents}
\author{Matthew I. Jones$^1$, Scott D. Pauls$^1$, Feng Fu$^{1,2}$}
\address{$^1$ Department of Mathematics, Dartmouth College, Hanover, NH 03755, USA\\
$^2$ Department of Biomedical Data Science, Geisel School of Medicine at Dartmouth, Lebanon, NH 03756, USA}

\begin{abstract}
Global coordination is required to solve a wide variety of challenging collective action problems from network colorings to the tragedy of the commons. Recent empirical study shows that the presence of a few noisy autonomous agents can greatly improve collective performance of humans in solving networked color coordination games. To provide further analytical insights into the role of behavioral randomness, here we study myopic artificial agents attempt to solve similar network coloring problems using decision update rules that are only based on local information but allow random choices at various stages of their heuristic reasonings. We consider that agents are distributed over a random bipartite network which is guaranteed to be solvable with two colors. Using agent-based simulations and theoretical analysis, we show that the resulting efficacy of resolving color conflicts is dependent on the specific implementation of random behavior of agents, including the fraction of noisy agents and at which decision stage noise is introduced. Moreover, behavioral randomness can be finely tuned to the specific underlying population structure such as network size and average network degree in order to produce advantageous results in finding collective coloring solutions. Our work demonstrates that distributed greedy optimization algorithms exploiting local information should be deployed in combination with occasional exploration via random choices in order to overcome local minima and achieve global coordination. 
\end{abstract}

\submitto{arXiv}
\maketitle

\section{Introduction}

Many classical games like the Prisoner's Dilemma focus on two players attempting to get the better of each other. Both players would like to defect while their opponent cooperates, thus reaping rewards and avoiding punishments. A great body of work is focused on how to foster cooperation in such non-zero sum games~\cite{nowak_2006_2,doebeli2005models}. 
But there is another well-studied class of games in which all players receive the most benefit when they work together, called coordination games~\cite{skyrms2004stag}. The optimal behavior for all players can be easily determined and agreed upon if all players can meet and strategize beforehand. In such games, the difficulty comes not from attempting to scam one's opponent, but figuring out what one's partner will play before choosing one's own strategy~\cite{huyck_battalio_beil_1990,nowak_2006}. However, there can still be a ``defecting'' component, in which one's opponent can unilaterally choose a strategy with lower maximum payoff but also less risk~\cite{Fang_Kimbrough_Pace_Valluri_Zheng_2002}.

Frequently, we consider playing games in which the population is given some spatial structure other than being well-mixed~\cite{durrett1994importance,szabo2005phase}. Population structure is typically modelled as a graph or network, where each node is an individual, and individuals play games if they are connected by an edge~\cite{ohtsuki2006simple,santos2005scale,fu_hauert_nowak_wang_2008,perc2010coevolutionary,rand2011dynamic,shirado2013quality,gomez2007dynamical,shirado2020network}. On such a network, many coordination games can be rephrased as network coloring problems~\cite{judd2010behavioral}. A coloring is a collection of labels or colors, one for each node, such that any two nodes connected by an edge have different colors. Network colorings make appearances in all sorts of fields, including sudoku puzzles, register allocation in computer science \cite{chaitin_1982}, and clustering problems \cite{hansen_delattre_1978}. Deciding on a time table for various classes with shared classrooms \cite{werra_1985} and assignment of radio frequencies \cite{zoeliner_beall_1977} are just two examples or coordination games that manifest naturally as network coloring problems. Generally, we let the nodes be individuals (which we refer to as artificial agents in this work), and the color choice represents the strategy of that individual. When the nodes of a network are properly colored, all the individuals are playing an optimal strategy. In this sense, the network coloring problem, if assigned with a proper payoff structure for the coloring outcome, can be considered broadly as coordination game~\cite{kun2013anti,apt2014coordination}.

In general, the network coloring problem is NP-hard \cite{garey_johnson_1999}. It is found that many difficult mathematical problems cannot be solved by a simple, direct approach, but it can help to apply a small degree of randomness to any algorithms searching the solution space. This approach has been used to all sorts of problems, including the Traveling Salesman Problem \cite{bonomi_lutton_1984} and the graph coloring problem \cite{johnson_aragon_mcgeoch_schevon_1991} with which we are concerned in this paper.

Attempts to solve the network coloring problem typically use information about the entire network to make decisions about the colors of nodes. This makes sense as having all the information simultaneously leads to better informed decisions. For example, Ref.\cite{johnson_aragon_mcgeoch_schevon_1991} uses a notion of temperature to gradually reduce stochastic behavior as the system ``cools" into the global solution. This requires some central information unit that instructs each node on color choice. However, if we are using the network as a model of a population in which edges represent interactions, such a central ``brain" may not exist. Instead, individuals may be forced to make decisions based on nothing except the color of their neighbors at any given moment. Thus, solving the distributed network coloring problem, in which each node decides its color with only the local information about its neighbors, is more difficult, as we lose the ability to make decisions based on the state of the entire network.

Some work has already been done in solving distributed coloring problems. One line of work involves deterministic algorithms that require more colors than necessary for the network \cite{finocchi_panconesi_silvestri_2004, chaudhuri_graham_jamall_2008}. The additional available colors make the problem much more tractable. There has also been work involving experiments with human subjects who have been given control of the color of a single node, and are asked to choose colors to eliminate conflicts with their neighbors. Ref.~\cite{kearns_2006} observed that individuals would frequently choose colors that temporarily increased the total number of color conflict, but ultimately lead to a global coloring. Ref.~\cite{shirado_christakis_2017} found that by adding a small number of bots (namely, artificial agents as opposed to humans) to the system who periodically made random changes ``decreased both the number of conflicts and the duration of unresolvable conflicts" when finding network colorings. However, they also found that the bots could be detrimental if not properly tuned with the appropriate levels of randomness.~\cite{shirado_christakis_2017}. 

Motivated by these empirical findings~\cite{kearns_2006,shirado_christakis_2017}, here we study artificial agents using stochastic decision update rules to solve the network coloring problem collectively. Without loss of generality, we assume agents are situated on the simple case of networks that can be colored with only two colors, often called bipartite networks~\cite{guillaume2006bipartite}. This specific network structure simplifies the number of possible colorings (exactly two for a connected network) and offers analytical insights that would be formidable to obtain otherwise. The results reported below come from an entire population of artificial agents (in the fashion of simulated bots as in Ref.~\cite{shirado_christakis_2017}), some of whom are behaving deterministically and some stochastically. Our work sheds some light on the appropriate levels of randomness to optimize solving the distributed coloring problem.

\section{Methods}

\subsection{Random Network Construction}

As we will see, different network topologies will be easier or harder to color. Even with global information, finding network colorings becomes exponentially more difficult as the number of nodes increases~\cite{garey_johnson_1999}. On the other hand, as average degree increases, individuals will have more neighbors and therefore be able to make more informed decisions when choosing a color. Throughout this paper, we simulate artificial agents that attempt to find 2-colorings of random bipartite networks. The exact structure of these networks will vary, as will the decision update rules agents use to solve the network colorings. 

We construct a random network with $n$ nodes and average degree $k$ by first assigning each node to group A or group B with probability $\frac{1}{2}$. Then, we add an edge between any two nodes in different groups with probability $\frac{2k}{n}$. Thus, the resulting network is guaranteed to have a 2-coloring by assigning every node in group A one color and every node in group B the other color. However, there may be different numbers of nodes for each color, as the sizes of groups A and B are binomially distributed in our bipartite network model.

\subsection{Decision Update Rules for Agents}
In this paper, we consider multiple decision update rules to account for a variety of artificial agents' behavior, each with their own strengths and weaknesses. In the following, an \emph{acceptable} local coloring at a node is the choice of color such that none of the node's neighbors have that color (no color conflicts with neighbors). 

We first consider a basic \emph{greedy} update rule of agents:

\paragraph{I: Basic greedy update rule}
\begin{enumerate}
    \item[Step 1:] Check if the current color is already an acceptable local coloring. If yes, no further change.
    \item[Step 2:] If not, check if the other color would make an acceptable local coloring. If yes, choose the other color.
    \item[Step 3:] (and so on for all the steps) If not, choose the color that minimizes color conflicts, and randomly choose one color if both color choices have the same number of color conflicts with neighbors. 
    \end{enumerate}

We incorporate random behavior in various decision stages in the following modified update rules based on the basic greedy update rule above.

\paragraph{II: Randomness-first update rule}

\begin{enumerate}
    \item[Step 1] With probability $p$, choose a color uniformly at random.
    \item[Step 2] Check if the chosen color is already an acceptable local coloring. If yes, no further change.
    \item[Step 3] If not, check if the other color would make an acceptable local coloring. If yes, choose the other color.
    \item[Step 4] If not, choose the color that minimizes color conflicts, and randomly choose one color if both color choices have the same number of color conflicts with neighbors. 
\end{enumerate}

\paragraph{III: Memory-$0$ update rule}

\begin{enumerate}
    \item[Step 1] Check if the current color is already an acceptable local coloring. If yes, no further change.
    \item[Step 2] If not, check if the other color would make an acceptable local coloring. If yes, choose the other color.
    \item[Step 3] If not, with probability $p$, choose a color uniformly at random.
    \item[Step 4] Choose the color of the two that minimizes color conflicts, and randomly choose one color if both color choices have the same number of color conflicts with neighbors. 
\end{enumerate}

\paragraph{IV: Memory-$N$ update rule}

\begin{enumerate}
    \item[Step 1] Check if the current color is already an acceptable local coloring. If yes, no further change.
    \item[Step 2] If not, check if the other color would make an acceptable local coloring. If yes, choose the other color.
    \item[Step 3] If not, and if no neighbors have changed colors in prior $N$ cycles, with probability $p$, choose a color uniformly at random.
    \item[Step 4] Choose the color of the two that minimizes color conflicts, and randomly choose one color if both color choices have the same number of color conflicts with neighbors. 
\end{enumerate}

\subsection{Initialization of Agent Behavior}
Each artificial agent, located at a node in the network, behaves according to one of the aforementioned update rules. Specifically, we consider scenarios where the population may be using two different update rules. A certain fraction $\rho_r$ of randomly-selected agents adopt one of the randomness-first, memory-$0$, or memory-$N$ update rules where the propensity of random behavior is $p$ (as defined in the update rules), and the rest of agents use the basic greedy update rule.

The color choice of agents is updated in a random sequential manner~\cite{szabo2007evolutionary}. Agents update one at a time, and the order in which agents update is random. Each agent begins with a randomly chosen color.

\subsection{Difficulty Metrics}

We use three different metrics to quantify how successful a given decision update rule is in solving coloring problems by artificial agents: the number of unsolved networks, the number of update cycles, and the number of player updates. 

The number of unsolved networks metric is simply the probability that a given network will reach a coloring given certain initial conditions including update order, update rules for each agent, and initial coloring. 

The number of update cycles measures the number of times each agent goes through the update process, and the number of updated agents measures the total number of color changes. Roughly, the number of update cycles measures how long it will take the system to reach a coloring in real time, and the number of updated agents measures how involved the process is for all agents involved. Because some combinations of networks and initial conditions may never end up with a complete coloring solution, these metrics have the possibility to be infinite in these cases. Therefore, the average of difficulty metrics across model parameter combinations may be heavily skewed by some of the unsolved network coloring cases. Nevertheless, these difficulty metrics provide a practical means to compare the efficacy of resolving color conflicts across simulated scenarios and can help reveal interesting results to some extent. 

\section{Results}

\subsection{Bowties and Gridlock}
\begin{figure}
    \centering
    \includegraphics[width=\textwidth]{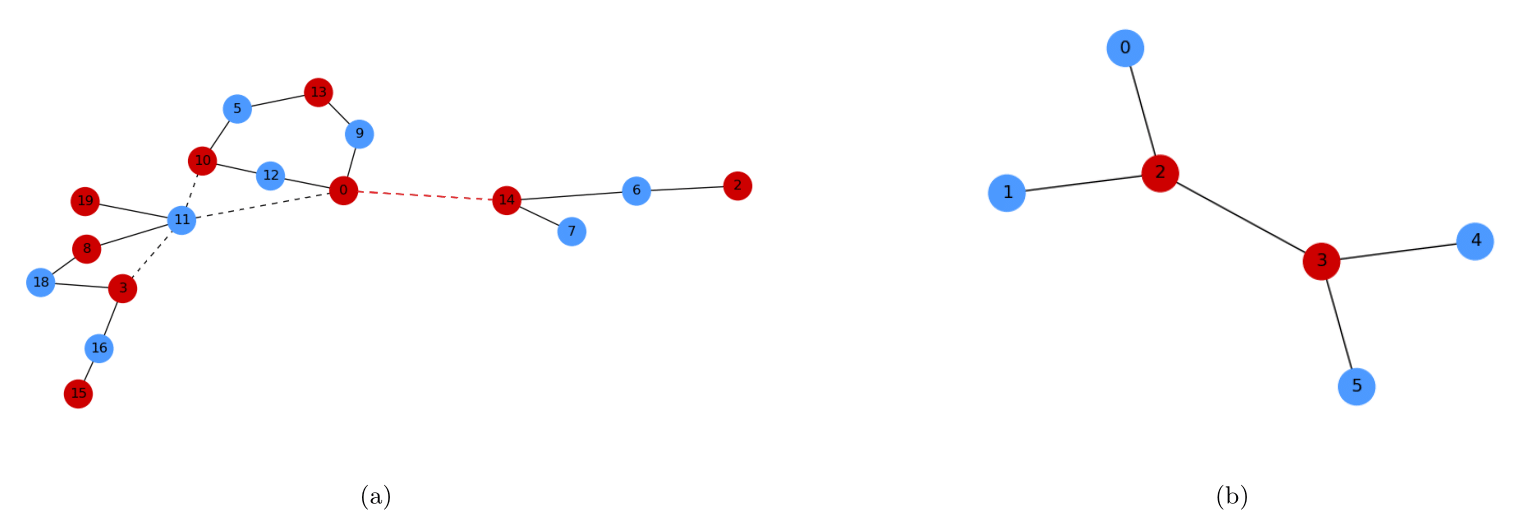}
    \caption{Overcoming local minima is often needed to solve collective action problems. (a) shows a small network that did not find a valid coloring using only greedy behavior. The four dashed edges represent ``bowties,'' subgraphs where the greedy update rule can become gridlocked. The red edge shows a color conflict that cannot be resolved by greedy behavior. In (b), we see how the interior nodes of a bowtie are both forced to keep the same color by the exterior nodes, creating gridlock.}
    \label{fig:bowtieFig}
\end{figure}

To see how local minima arise, we show a small network in which each agent occupying a network node uses the greedy update rule in Fig.~\ref{fig:bowtieFig}a. The dashed edges are ``bowties,'' small subgraphs consisting of a central edge whose end nodes both have at least three edges. Motif structures like this can lead to gridlock and the failure of the greedy update rule, as demonstrated in Fig.~\ref{fig:bowtieFig}b. If the central agents are playing the same color, they can become locked in by their other neighbors, and as a consequence, the greedy update rule becomes trapped at this local minimum, unable to explore the entire space and find a global minimum of color conflicts. Without random behavior, the network will never reach a global coloring once this happens. The smallest possible network structure that can become gridlocked is the six-node bowtie, as shown in Fig.~\ref{fig:bowtieFig}b.

This simple case demonstrated in Fig.~\ref{fig:bowtieFig}b can yield an interesting insight. Consider the case where there is no random behavior and each agent is playing the greedy update rule. There are $6!\cdot 2^6$ random initial conditions for the update order and initial colors. Using exhaustive search to work out each case, we find that the simple bowtie results in gridlock with probability $\frac{29}{120}$. In each case, either gridlock or a global coloring is always reached after two update cycles. 

Of course, brute-force computation quickly becomes untenable for large network sizes, but we can still develop helpful intuition from this simple example (Fig.~\ref{fig:bowtieFig}b). With the randomness-first update rule, if at least one agent has random behavior (occurs with probability $1-(1-\rho_r)^6$), the network will eventually find a global coloring. However, in the memory-$N$ update rule, the peripheral nodes already have a locally acceptable color, and will not change even if they have the potential for random behavior. One of the middle two nodes must have random behavior to find a coloring, which happens with probability $1-(1-\rho_r)^2$, a much less likely event than in the randomness-first update rule. Thus, the gridlock probabilities for the randomness-first and memory-$N$ update rules respectively are approximately

\begin{equation}\label{eq:rFGridlock}
P_{\textrm{rand-first}}(\textrm{Gridlock})=\frac{29}{120} (1-\rho_r)^6\end{equation}
\begin{equation}\label{eq:m0Gridlock}
P_{\textrm{memory-$N$}}(\textrm{Gridlock})=\frac{29}{120} (1-\rho_r)^2\end{equation}

\begin{figure}
    \centering
    \includegraphics[width = \textwidth]{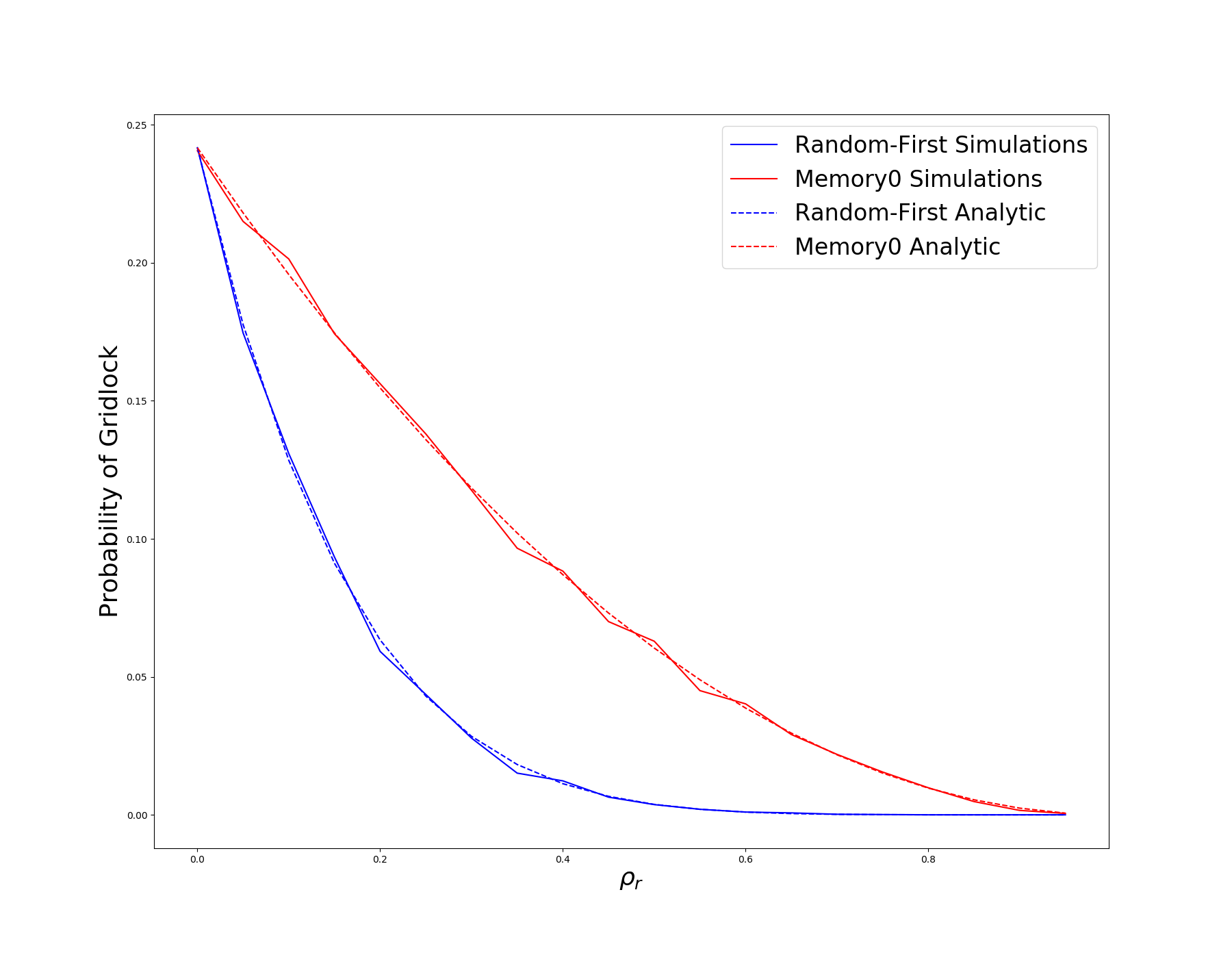}
    \caption{The probability of gridlock in the six-node bowtie for varying the fraction of agents with random behavior, $\rho_r$. We see that the simulations (using $p=0.5$) match well with the analytic results in Eqs.~\ref{eq:rFGridlock} and \ref{eq:m0Gridlock}. Here we compare the randomness-first rule with the memory-$0$ rule. Simulation results are averaged over $1,000$ independent runs.}
    \label{fig:bowtieSimPlot}
\end{figure}

We see excellent agreement between these equations and simulations in Fig.~\ref{fig:bowtieSimPlot}. We note that these probabilities are less accurate when $p$ is large, because individuals could behave randomly before the system reaches gridlock, disrupting the earlier computation for $\frac{29}{120}$ which assumed no random behavior takes place in the first two update cycles.

Similarly, we see that the memory-$N$ update rules require larger $\rho_r$ than the randomness-first rule to reach the same efficacy of resolving color conflicts. When using the former update rule, only agents with a color conflict are allowed to make random choices, unlike the latter randomness-first update rule. Because random behavior is limited to individuals with a color conflict, large $\rho_r$ values are less likely to result in too much randomness when most agents are already in a local coloring without conflicts and hence will not behave randomly in any given time step. We shall see this difference between randomness-first and memory-$N$ update rules manifest itself in simulations on larger networks in the following section.

\subsection{Monte Carlo Agent-Based Simulations}

\begin{figure}
    \centering
    \includegraphics[width = \textwidth]{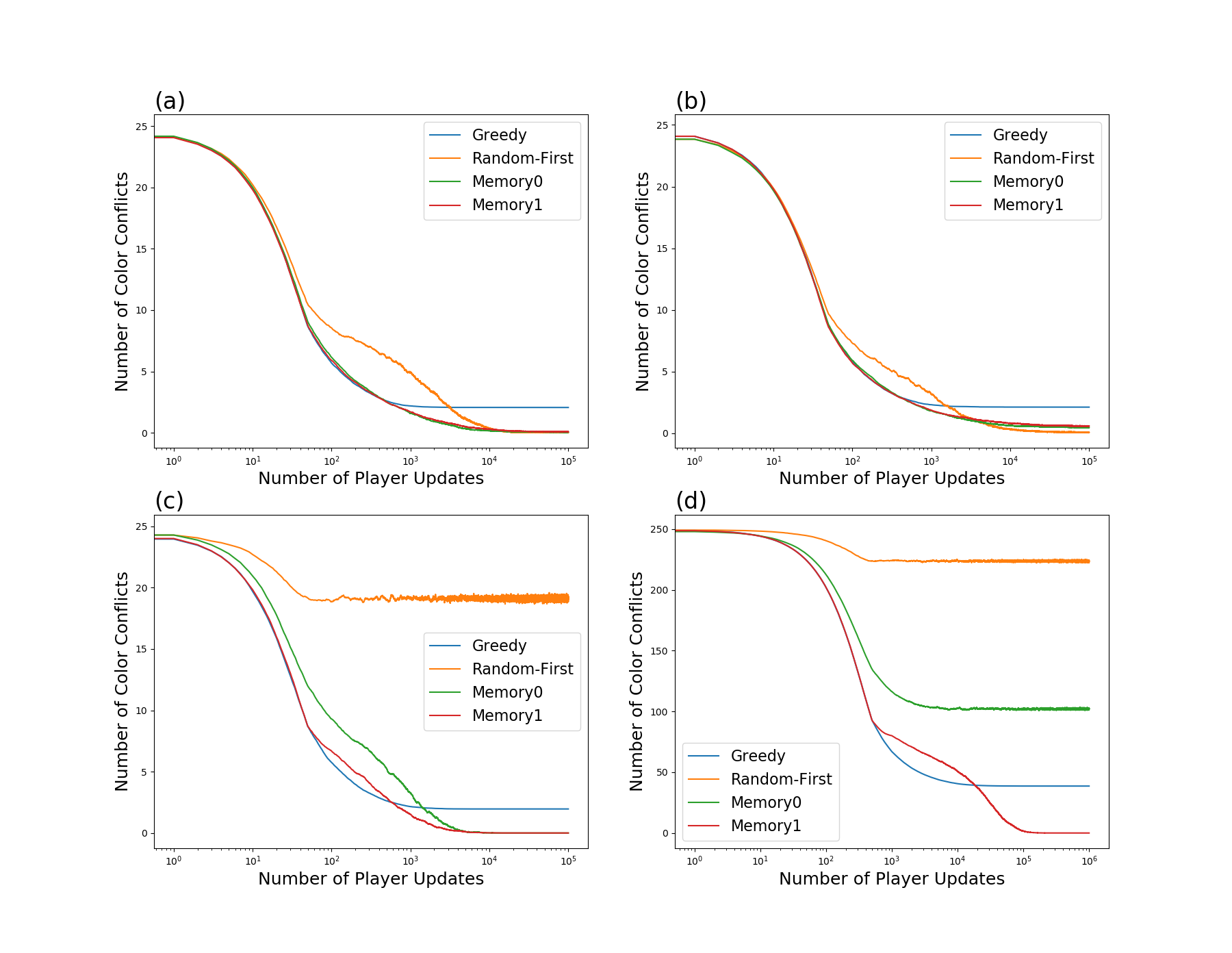}
    \caption{Plots of total conflicts vs time. Each curve is the average of $1,000$ simulations, and each run consists of $2,000$ update cycles. Observe that the x-axis is log-scale, to show the short and long term behavior of each update rule. All networks have average degree $2$, and the other network properties are as follows: a) $n=50, p=0.1, \rho_r = 0.9$ b) $n=50, p=0.1, \rho_r = 0.5$ c) $n=50, p=0.6, \rho_r = 1$ d) $n=500, p=0.6, \rho_r=1$}
    \label{fig:conflictCounts}
\end{figure}

Having defined the model parameters for the problem, we now can ask a basic question: What is the optimal amount of randomness to have in the system so as to reach a coloring solution? It turns out that the answer varies, depending on the specific update rule used, the size of the network, and the average degree of the underlying network. Typically, we will consider large and small networks with $50$ and $500$ nodes, and vary with average network degree values of 2 and 20, respectively. Figure \ref{fig:conflictCounts} shows how noisy agents using different update rules succeed at reducing the total number of conflicts in different situations. Notice that no update rule alone can beat the greedy update rule in the short term, but eventually the randomness-based update rules begin under-performing the greedy rule only to eventually surpass it and completely eliminate color conflicts.

There are two sources of difficulty for coloring networks using any randomness-based update rule. If there is not enough randomness, the decision update rule is unable to break away from the local minimum found by agents using the greedy update rule. If there is too much randomness, the probability that at least one agent will be picking the wrong color every turn is so high that the network will not find a coloring in a reasonable number of time steps. Methods like simulated annealing avoid this problem by cooling the system and decreasing the amount of randomness over time~\cite{johnson_aragon_mcgeoch_schevon_1991}. However, in a distributed system (where each agent is using only local information to choose color) with no global information like temperature, we are limited to very simple local update rules that simply cannot evolve over time.

\subsection{Randomness-first rule}
\begin{figure}
    \centering
    \includegraphics[width = \textwidth]{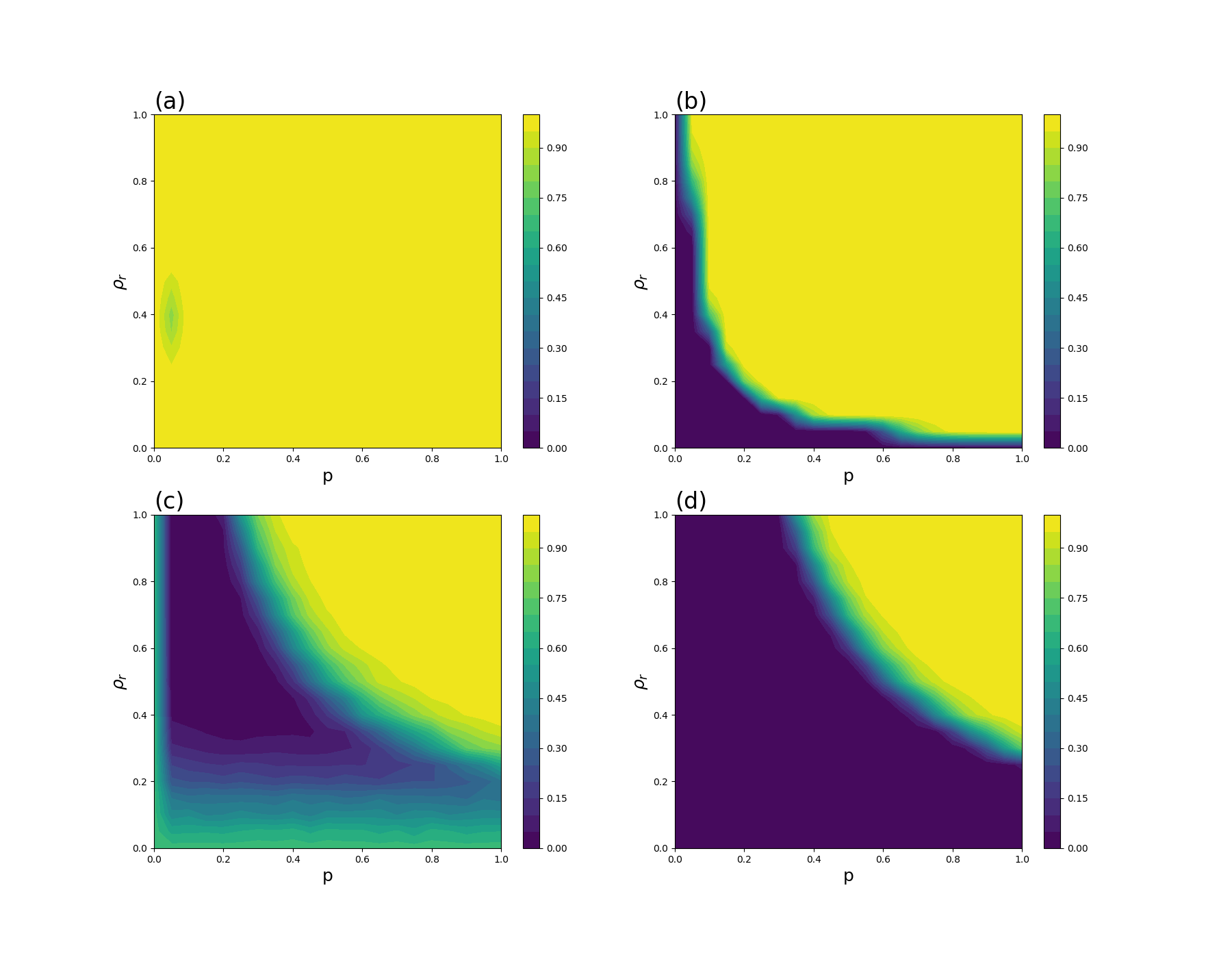}
    \caption{For the randomness-first update rule, simulation results of the probability of not solving the network in $10,000$ time steps using four different types of networks as a function of the level of randomness $p$ and the fraction of agents with random behavior $\rho_r$. The bipartite network parameters including the size $N$ and the average degree $k$ used for the underlying networks are as follows: a) $n=500,k=2$ b) $n=500,k=20$ c) $n=50,k=2$, d) $n=50,k=20$.}
    \label{fig:rFPlots}
\end{figure}

For the randomness-first update rule, we ran simulations for $20$ combination values of $\rho_r$ and $p$ between $0$ and $1$. Networks that found a coloring within $10,000$ update cycles by agents were considered solved, and those that did not find a coloring within $10,000$ cycles were considered unsolved. In Fig.~\ref{fig:rFPlots}, we show the results of these simulations.

We see the difficulty of too much and too little randomness in Fig.~\ref{fig:rFPlots}. In all four regions of the network parameter space (small/large size, low/high edge density) the probability of solving the network goes to zero because agents are always making random decisions, even when the rest of the network has found a local coloring. When average degree is two, we also see unsolved networks when there is very little randomness. Here, there are too few random agents to break out of the local minimum.

These results demonstrate how the randomness-first update rule's success varies depending on the properties of the network (Fig.~\ref{fig:rFPlots}). When average degree is high, randomness is actually a hindrance; the fewer random actions there are, the better. However, when average degree is low, a large fraction of the population using the randomness-first update rule with a low $p$ is best. Unfortunately, for large networks with small average degree, there seems to be no good $p$ and $\rho_r$ when using the randomness-first rule.

Notice that in general, as network size goes up and/or average degree goes down, there are more unsolved networks. This makes intuitive sense, as additional nodes means more colors that need to be correct, and smaller average degree means the nodes have less information and make poorer decisions. 

\subsection{Memory-$N$ rule}
\begin{figure}
    \centering
    \includegraphics[width = \textwidth]{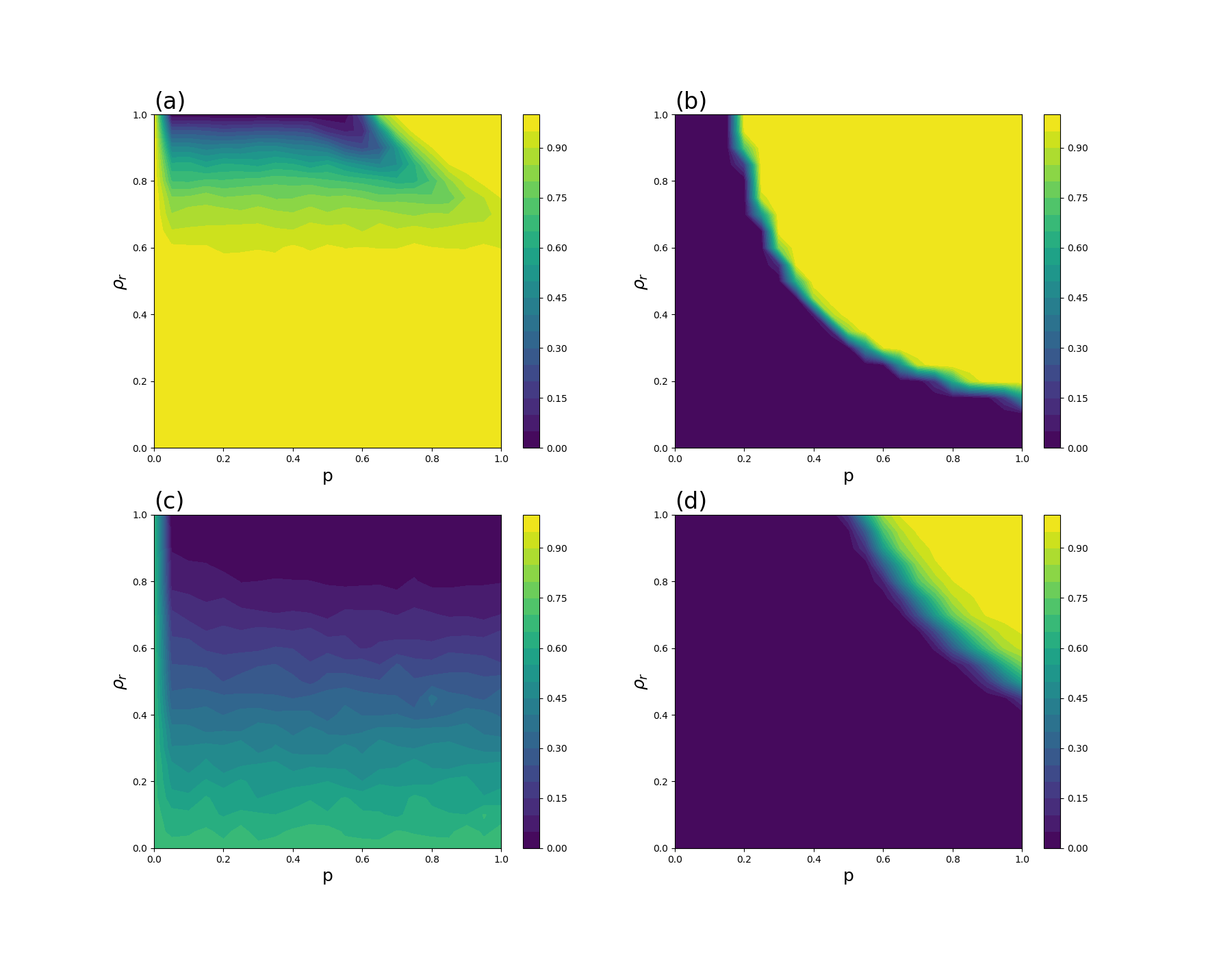}
    \caption{For the memory-$0$ update rule, simulation results of the probability of not solving the network in $10,000$ time steps using four different types of networks as a function of the level of randomness $p$ and the fraction of agents with random behavior $\rho_r$. The bipartite network parameters including the size $N$ and the average degree $k$ used for the underlying networks are as follows: a) $n=500,k=2$ b) $n=500,k=20$ c) $n=50,k=2$, d) $n=50,k=20$.}
    \label{fig:m0Plots}
\end{figure}

We first study the memory-$0$ update rule that differs from the randomness-first rule in that agents only take random actions if they are in conflict with at least one of their neighbors. Thus, there are fewer needless random actions, and we would expect this decision update rule to perform better where excess randomness is an issue. This is partially confirmed by simulations in Fig.~\ref{fig:m0Plots}.

\begin{figure}
    \centering
    \includegraphics[width = \textwidth]{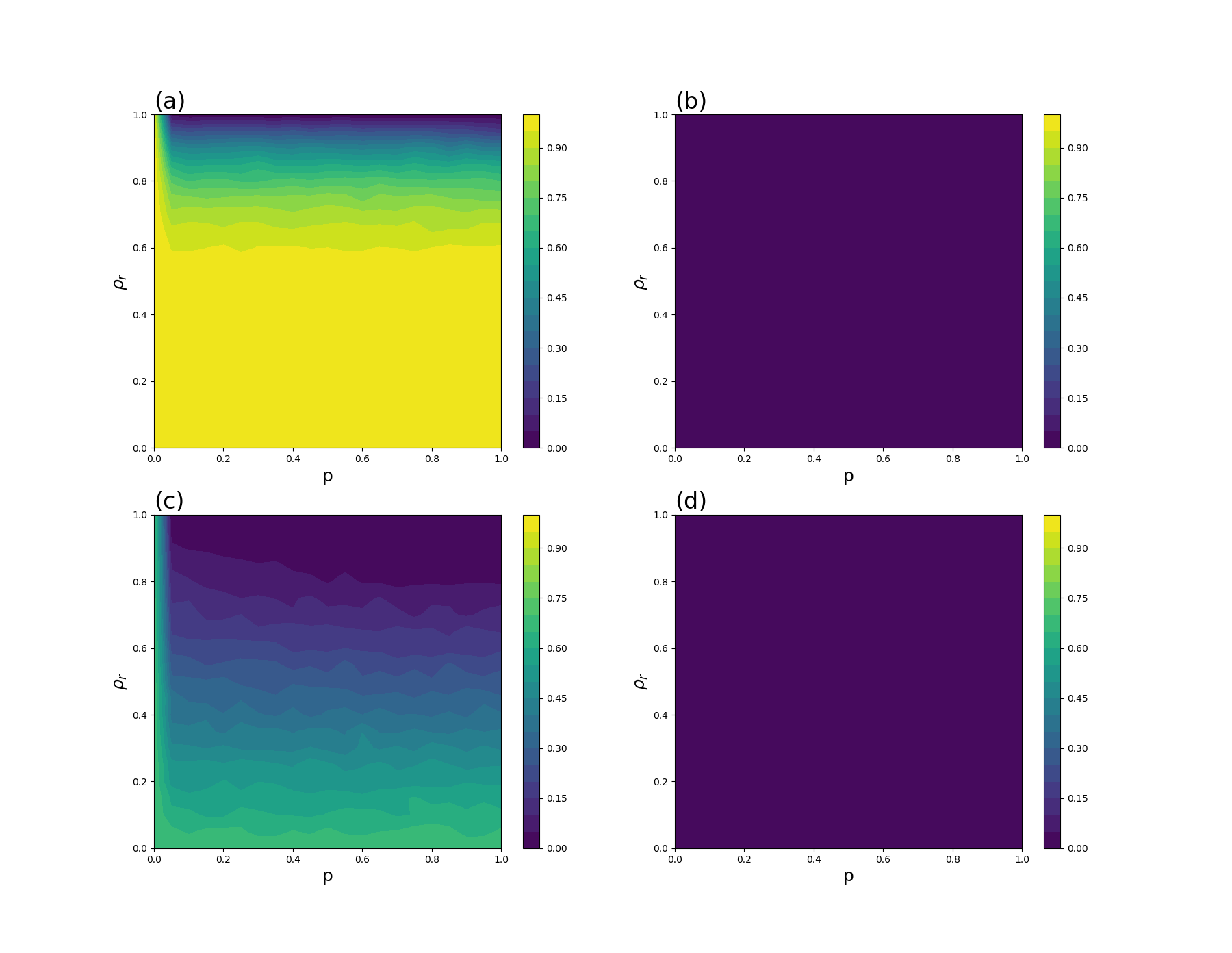}
    \caption{For the memory-$1$ update rule, simulation results of the probability of not solving the network in 10,000 time steps using four different types of networks as a function of the level of randomness $p$ and the fraction of agents with random behavior $\rho_r$. The bipartite network parameters including the size $N$ and the average degree $k$ used for the underlying networks are as follows: a) $n=500,k=2$ b) $n=500,k=20$ c) $n=50,k=2$, d) $n=50,k=20$.}
    \label{fig:m1Plots}
\end{figure}

Generally, we see an improvement of performance over the randomness-first update rule. The memory-$0$ rule does very well when $\rho_r$ is close to one, even for large networks with low average degree. However, it still struggles with excess randomness, particularly when network size and average degree are large. A higher average degree means that a single random color choice creates more color conflicts and therefore makes it more difficult for the system to settle into a global coloring. However, if we assume agents with a longer memory (i.e., $N\ge1$), this issue vanishes, as demonstrated in Fig.~\ref{fig:m1Plots}.

This compelling evidence suggests that the memory-$1$ update rule is most effective at resolving color conflicts as compared with the basic greedy update rule and the memory-$0$ update rule (cf. Figs.~\ref{fig:rFPlots}, \ref{fig:m1Plots}, and \ref{fig:m0Plots}). If $\rho_r$ is close to one, networks are almost always able to find a global coloring, regardless of network size or average degree. However, if for some reason only a rather small fraction of the agents are allowed to use randomness-based update rules, the randomness-first update rule will have more success, as seen in the simple bowtie example in Fig.~\ref{fig:bowtieFig}b.

\section{Discussion \& Conclusion}

Among others, an important insight stemming from the present paper is that the type of decision update rule used by agents is at least as important as the amount of random behavior. The randomness-first and memory-$N$ update rules require different conditions to be successful. This gives us two different update rules that are useful in different settings, and should be thought of as complementary instead of one being superior to the other. For example, in a scenario where all agents are able to use a randomness-based update rule, a memory-$N$ update rule can be used to great success. However, if only a few agents in the population can be persuaded to take on the personal risk of behaving randomly (or a small number of bots prescribed with random behavior have been introduced into the population like \cite{shirado_christakis_2017}), a randomness-first update rule with a low $p$ will have a higher chance of success.

This paper most closely relates to previous work involving human subjects playing the coloring game with random bots~\cite{shirado_christakis_2017}. While random behavior was observed coming from human players~\cite{kearns_2006}, it is not clear if this behavior was closer to the randomness-first or the memory-$N$ update rule. The noisy bots themselves in Ref.~\cite{shirado_christakis_2017} played a randomness-first update rule, which may explain how such a small fraction ($\rho_r = 0.15$) of random actors had such a profound impact on the network coloring game.

Our work demonstrates that the solving of challenging distributed network coloring problems can be achieved by entirely using myopic artificial agents without human subjects. We find that it is necessary to have enough randomness to ensure that the system is able to find the global coloring, but without having so much random behavior the system never settles down. Different randomness-based update rules each are successful in different networks with different parameters. 

Of particular note, here we only consider the simplest possible 2-colorings of bipartite networks, which is surely an over-simplification of the more general coloring problems. Introducing even one more color adds all sorts of complications. For example, the bowtie analysis completely falls apart, as the subgraphs to result in gridlock in a 3-colorable network are significantly larger and more complex. Besides, this paper also only considers populations that play a mix of two decision update rules: a fraction of the agents use greedy decision rule and the rest use randomness-based rule. It is possible that other potential combinations, such as a mixed population of agents using the randomness-first rule and the memory-$N$ rule, could succeed in places where neither update rule succeeds alone. Future work taking into account these extensions will be of interest and improves our understanding of collective decision-makings in the presence of noise~\cite{couzin2011uninformed,couzin2005effective}, and more generally, machine behavior~\cite{rahwan2019machine}.

\section{Acknowledgements}
F.F. is supported by the Bill \& Melinda Gates Foundation (award no. OPP1217336), the NIH COBRE Program (grant no. 1P20GM130454), a Neukom CompX Faculty Grant, the Dartmouth Faculty Startup Fund and the Walter \& Constance Burke Research Initiation Award.


\section{References}
\begin{thebibliography}{10}

\bibitem{nowak_2006_2}
M.~A. Nowak.
\newblock Five rules for the evolution of cooperation.
\newblock {\em Science}, 314(5805):1560?1563, 2006.

\bibitem{doebeli2005models}
Michael Doebeli and Christoph Hauert.
\newblock Models of cooperation based on the prisoner's dilemma and the
  snowdrift game.
\newblock {\em Ecology letters}, 8(7):748--766, 2005.

\bibitem{skyrms2004stag}
Brian Skyrms.
\newblock {\em The stag hunt and the evolution of social structure}.
\newblock Cambridge University Press, 2004.

\bibitem{huyck_battalio_beil_1990}
John B.~Van Huyck, Raymond~C. Battalio, and Richard~O. Beil.
\newblock Tacit coordination games, strategic uncertainty, and coordination
  failure.
\newblock {\em The American Economic Review}, 80(1):234--248, 1990.

\bibitem{nowak_2006}
Martin~A. Nowak.
\newblock {\em Evolutionary dynamics: exploring the equations of life}.
\newblock Belknap Press of Harvard University Press, 2006.

\bibitem{Fang_Kimbrough_Pace_Valluri_Zheng_2002}
Christina Fang, Steven~O. Kimbrough, Stefano Pace, Annapurna Valluri, and
  Zhiqiang Zheng.
\newblock On adaptive emergence of trust behavior in the game of stag hunt.
\newblock {\em Group Decision and Negotiation}, 11(6):449--467, 2002.

\bibitem{durrett1994importance}
Richard Durrett and Simon Levin.
\newblock The importance of being discrete (and spatial).
\newblock {\em Theoretical population biology}, 46(3):363--394, 1994.

\bibitem{szabo2005phase}
Gy{\"o}rgy Szab{\'o}, Jeromos Vukov, and Attila Szolnoki.
\newblock Phase diagrams for an evolutionary prisoner?s dilemma game on
  two-dimensional lattices.
\newblock {\em Physical Review E}, 72(4):047107, 2005.

\bibitem{ohtsuki2006simple}
Hisashi Ohtsuki, Christoph Hauert, Erez Lieberman, and Martin~A Nowak.
\newblock A simple rule for the evolution of cooperation on graphs and social
  networks.
\newblock {\em Nature}, 441(7092):502--505, 2006.

\bibitem{santos2005scale}
Francisco~C Santos and Jorge~M Pacheco.
\newblock Scale-free networks provide a unifying framework for the emergence of
  cooperation.
\newblock {\em Physical review letters}, 95(9):098104, 2005.

\bibitem{fu_hauert_nowak_wang_2008}
Feng Fu, Christoph Hauert, Martin~A. Nowak, and Long Wang.
\newblock Reputation-based partner choice promotes cooperation in social
  networks.
\newblock {\em Physical Review E}, 78(2), 2008.

\bibitem{perc2010coevolutionary}
Matja{\v{z}} Perc and Attila Szolnoki.
\newblock Coevolutionary games?a mini review.
\newblock {\em BioSystems}, 99(2):109--125, 2010.

\bibitem{rand2011dynamic}
David~G Rand, Samuel Arbesman, and Nicholas~A Christakis.
\newblock Dynamic social networks promote cooperation in experiments with
  humans.
\newblock {\em Proceedings of the National Academy of Sciences},
  108(48):19193--19198, 2011.

\bibitem{shirado2013quality}
Hirokazu Shirado, Feng Fu, James~H Fowler, and Nicholas~A Christakis.
\newblock Quality versus quantity of social ties in experimental cooperative
  networks.
\newblock {\em Nature communications}, 4(1):1--8, 2013.

\bibitem{gomez2007dynamical}
Jes{\'u}s G{\'o}mez-Gardenes, Michel Campillo, Luis~Mario Flor{\'\i}a, and
  Yamir Moreno.
\newblock Dynamical organization of cooperation in complex topologies.
\newblock {\em Physical Review Letters}, 98(10):108103, 2007.

\bibitem{shirado2020network}
Hirokazu Shirado and Nicholas~A Christakis.
\newblock Network engineering using autonomous agents increases cooperation in
  human groups.
\newblock {\em Iscience}, 23(9):101438, 2020.

\bibitem{judd2010behavioral}
Stephen Judd, Michael Kearns, and Yevgeniy Vorobeychik.
\newblock Behavioral dynamics and influence in networked coloring and
  consensus.
\newblock {\em Proceedings of the National Academy of Sciences},
  107(34):14978--14982, 2010.

\bibitem{chaitin_1982}
G.~J. Chaitin.
\newblock Register allocation \& spilling via graph coloring.
\newblock {\em ACM SIGPLAN Notices}, 17(6):98?101, 1982.

\bibitem{hansen_delattre_1978}
Pierre Hansen and Michel Delattre.
\newblock Complete-link cluster analysis by graph coloring.
\newblock {\em Journal of the American Statistical Association},
  73(362):397?403, 1978.

\bibitem{werra_1985}
D.~De Werra.
\newblock An introduction to timetabling.
\newblock {\em European Journal of Operational Research}, 19(2):151?162,
  1985.

\bibitem{zoeliner_beall_1977}
J.~Zoeliner and C.~Beall.
\newblock A breakthrough in spectrum conserving frequency assignment
  technology.
\newblock {\em IEEE Transactions on Electromagnetic Compatibility},
  EMC-19(3):313?319, 1977.

\bibitem{kun2013anti}
Jeremy Kun, Brian Powers, and Lev Reyzin.
\newblock Anti-coordination games and stable graph colorings.
\newblock In {\em International Symposium on Algorithmic Game Theory}, pages
  122--133. Springer, 2013.

\bibitem{apt2014coordination}
Krzysztof~R Apt, Mona Rahn, Guido Sch{\"a}fer, and Sunil Simon.
\newblock Coordination games on graphs.
\newblock In {\em International Conference on Web and Internet Economics},
  pages 441--446. Springer, 2014.

\bibitem{garey_johnson_1999}
Michael~R. Garey and David~S. Johnson.
\newblock {\em Computers and intractability}.
\newblock Freeman, 1999.

\bibitem{bonomi_lutton_1984}
Ernesto Bonomi and Jean-Luc Lutton.
\newblock The n-city travelling salesman problem: Statistical mechanics and the
  metropolis algorithm.
\newblock {\em SIAM Review}, 26(4):551?568, 1984.

\bibitem{johnson_aragon_mcgeoch_schevon_1991}
David~S. Johnson, Cecilia~R. Aragon, Lyle~A. Mcgeoch, and Catherine Schevon.
\newblock Optimization by simulated annealing: An experimental evaluation; part
  ii, graph coloring and number partitioning.
\newblock {\em Operations Research}, 39(3):378?406, 1991.

\bibitem{finocchi_panconesi_silvestri_2004}
Irene Finocchi, Alessandro Panconesi, and Riccardo Silvestri.
\newblock An experimental analysis of simple, distributed vertex coloring
  algorithms.
\newblock {\em Algorithmica}, 41(1):1?23, 2004.

\bibitem{chaudhuri_graham_jamall_2008}
Kamalika Chaudhuri, Fan~Chung Graham, and Mohammad~Shoaib Jamall.
\newblock A network coloring game.
\newblock {\em Lecture Notes in Computer Science Internet and Network
  Economics}, page 522?530, 2008.

\bibitem{kearns_2006}
M.~Kearns.
\newblock An experimental study of the coloring problem on human subject
  networks.
\newblock {\em Science}, 313(5788):824?827, 2006.

\bibitem{shirado_christakis_2017}
Hirokazu Shirado and Nicholas~A. Christakis.
\newblock Locally noisy autonomous agents improve global human coordination in
  network experiments.
\newblock {\em Nature}, 545(7654):370?374, 2017.

\bibitem{guillaume2006bipartite}
Jean-Loup Guillaume and Matthieu Latapy.
\newblock Bipartite graphs as models of complex networks.
\newblock {\em Physica A: Statistical Mechanics and its Applications},
  371(2):795--813, 2006.

\bibitem{szabo2007evolutionary}
Gy{\"o}rgy Szab{\'o} and Gabor Fath.
\newblock Evolutionary games on graphs.
\newblock {\em Physics reports}, 446(4-6):97--216, 2007.

\bibitem{couzin2011uninformed}
Iain~D Couzin, Christos~C Ioannou, G{\"u}ven Demirel, Thilo Gross, Colin~J
  Torney, Andrew Hartnett, Larissa Conradt, Simon~A Levin, and Naomi~E Leonard.
\newblock Uninformed individuals promote democratic consensus in animal groups.
\newblock {\em science}, 334(6062):1578--1580, 2011.

\bibitem{couzin2005effective}
Iain~D Couzin, Jens Krause, Nigel~R Franks, and Simon~A Levin.
\newblock Effective leadership and decision-making in animal groups on the
  move.
\newblock {\em Nature}, 433(7025):513--516, 2005.

\bibitem{rahwan2019machine}
Iyad Rahwan, Manuel Cebrian, Nick Obradovich, Josh Bongard, Jean-Fran{\c{c}}ois
  Bonnefon, Cynthia Breazeal, Jacob~W Crandall, Nicholas~A Christakis, Iain~D
  Couzin, Matthew~O Jackson, et~al.
\newblock Machine behaviour.
\newblock {\em Nature}, 568(7753):477--486, 2019.

\end{thebibliography}
\end{document}